\documentclass[a4paper,aps,prd,onecolumn,preprintnumbers,nofootinbib,11pt,twoside]{revtex4}

\pdfoutput=1
\usepackage[colorlinks=true,pdfstartview=FitV, linkcolor=red,citecolor=blue,urlcolor=magenta]{hyperref}

\usepackage{graphicx}
\usepackage{microtype}
\usepackage{amsmath}
\usepackage{amssymb}
\usepackage{subfigure}
\usepackage{hyperref}
\usepackage{url}
\usepackage{xcolor}
\usepackage{color}
\usepackage{mathrsfs}
\usepackage{calrsfs}
\usepackage{amsfonts}
\usepackage{eufrak}
\usepackage{tabularx}
\usepackage{eucal}
\usepackage{latexsym}
\usepackage{ragged2e}
\usepackage{epsfig}
\usepackage{textcomp}
\usepackage{subfigure}
\usepackage{marvosym}
\usepackage{ifsym}
\usepackage{lipsum}
\usepackage{appendix}
\usepackage{bm}
\usepackage{mathrsfs}

\begin{document}

\thispagestyle{empty}

\begin{center}


\title{Thermodynamic curvature in phase transitions for Hayward AdS black hole}

\author{Haximjan Abdusattar}
\email{axim@ksu.edu.cn}
\affiliation{School of Physics and Electrical Engineering, Kashi University, Kashi 844000, Xinjiang, China}


\begin{abstract}


We investigate $P$-$V$ criticality, with a focus on Ruppeiner geometry, in the extended phase space of Hayward anti-de Sitter (AdS) black holes. Through thermodynamic analysis, we confirm that Hayward-AdS black holes undergo distinct $P$-$V$ phase transitions and exhibit well-defined critical phenomena in the vicinity of their critical points. These behaviors are characterized by four critical exponents that typically obey the scaling laws predicted by mean-field theory--indicating a consistent thermodynamic framework with classical phase transition systems ($e.g.,$ van der Waals fluids).
Furthermore, we employ Ruppeiner geometry to probe the thermodynamic fluctuations of Hayward-AdS black holes, and by calculating the corresponding curvature scalar, we gain direct insights into the interaction nature of the black hole's microscopic constituents.

%

\end{abstract}

\maketitle
\end{center}

\section{Introduction}

Pioneering studies by Hawking and Bekenstein \citep{Bekenstein1972,Bardeen1973,Hawking:1976de} revealed the fundamental nature of black holes--not merely gravitational systems, but entities with distinct thermodynamic properties. Specifically, a black hole's temperature is determined by its surface gravity, while its entropy is directly associated with the area of its event horizon \citep{Bekenstein1973}. Crucially, the Hawking temperature of a black hole changes as it accretes matter or emits radiation \citep{Hawking:1974sw}; this thermodynamic behavior strongly implies that black holes must possess a microscopic structure \cite{Strominger:1996sh}. Since the prediction of this property, one of the most compelling questions in black hole physics has persisted: what exactly constitutes the microscopic state of a black hole?

Over the past few decades, the field of black hole thermodynamics has witnessed numerous groundbreaking discoveries. Similar to ordinary thermodynamic systems, black holes exhibit a rich variety of phase transitions--among which, those of black holes in anti-de Sitter (AdS) spacetimes have emerged as a focal point of recent research.
A core assumption underpinning these studies is the interpretation of the cosmological constant $\Lambda$ as a thermodynamic pressure variable \citep{Kastor:2009wy,Dolan:2010ha,Kubiznak:2016qmn}, defined as
\begin{equation}\label{PLambda}
P=-\frac{\Lambda}{8\pi G_{\rm N}}\,, \,\,\,\,\,\,\,\,\, \Lambda=-\frac{(D-1)(D-2)}{2 L^2} \,,
\end{equation}
where $L$ denotes the AdS radius and $G_{\rm N}$ is Newton's gravitational constant.
Within the framework of this definition, a variety of distinct phase transitions emerge in black hole thermodynamics, such as van der Waals \cite{Kubiznak:2012wp}, (multiple) re-entrant \cite{Altamirano:2013ane, Frassino:2014pha, Dehyadegari:2017hvd, Hendi:2017mfu}, those with isolated critical points \cite{Dolan:2014vba}, those with triple points \cite{Altamirano:2013uqa}, and superfluid-like phase transitions \cite{Hennigar:2016xwd}. (See Refs.\cite{Bhattacharya:2017hfj,Spallucci:2013osa,Majhi:2016txt, Altamirano:2014tva,Cai:2013qga,Hu:2018qsy} for more details on related works and extended analyses of these phase transitions.)\footnote{Notably, recent studies have revealed that scaling laws in black hole systems can be violated in four dimensions \cite{Hu:2024ldp} and higher dimensions \cite{Ahmed:2022kyv}. For the former case, thermodynamic curvature in phase transitions has also been investigated to probe the underlying microstructural properties \cite{Abdusattar:2024zzi}.}


By analyzing the phase transitions of black holes, thermodynamic geometry has emerged as a powerful tool for probing their microscopic structure, with Ruppeiner geometry playing a particularly pivotal role \citep{Ruppeinerb2008}. The core idea of this approach is to define a line element in the thermodynamic parameter space, which quantifies the ``distance" between two adjacent fluctuation states \citep{Ruppeiner95}. The curvature scalar constructed from this line element directly indicates the nature of interactions among the system's components: a positive curvature corresponds to repulsive interactions, while a negative curvature signifies attractive interactions. Notably, this interpretation is not unique to black holes, but is derived from the application of Ruppeiner geometry to ordinary thermodynamic systems \citep{Janyszek1990,Oshima1999}, whose cross-system consistency lends reliability to microscopic studies of black holes. Inspired by this, the microscopic origin of black hole thermodynamics has been explored \cite{Wei:2012ui,Wei:2015iwa,Wei:2019uqg,Wei:2019yvs,Guo:2019oad,Miao:2017fqg,KordZangeneh:2017lgs,Wei:2019ctz,Dehyadegari:2016nkd,Hu:2020pmr,Abdusattar:2023xxs}. It is important to note that the statistical mechanical study of black holes differs fundamentally from that of ordinary thermodynamic systems: while ordinary systems typically derive macroscopic thermodynamic quantities from the behavior of microscopic particles, black hole research requires ``reverse deduction"--inferring information about microscopic structures from known thermodynamic quantities by analyzing macroscopic phase transition behavior.

This paper focuses on the Hayward-AdS black hole, investigating its microstructure through a $P$-$V$ criticality analysis. A well-recognized limitation of classical gravity theories is the existence of spacetime singularities; quantum effects are widely believed to remedy this defect, highlighting the necessity of developing a quantum theory of gravity. However, no mature quantum gravity theory currently exists, making regular black hole models within a semiclassical framework an important avenue for studying singularity correction. Bardeen was the first to propose a singularity free regular black hole model \citep{bardeen1968non}, which replaces the singularity of traditional black holes with a de Sitter core. Subsequent studies further demonstrated that the charged version of the Bardeen regular black hole can be derived from Einstein gravity coupled to a nonlinear electrodynamic source \citep{AyonBeato:1998ub,AyonBeato:2000zs}, and can even be interpreted as the gravitational field generated by a nonlinear magnetic monopole. Hayward proposed another representative class of regular black hole solutions \citep{Hayward:2005gi}: similar to the Bardeen solution, it is a degenerate configuration of the gravitational field of a nonlinear source, carrying a magnetic charge and a free integration constant. In theories with infinite towers of higher-order curvature corrections in dimensions $D \geq 5 $, regular black holes with anti-de Sitter asymptotics are constructed \cite{Bueno:2024eig,Bueno:2024zsx,Hennigar:2025yqm};
the regularity of these black holes, along with related branes and solutions coupled to various electrodynamics, has been analyzed. Ref. \cite{Hennigar:2025yqm} also investigates the thermodynamics of such AdS black holes, revealing thermodynamic features associated with regularization parameters. Motivated by these works, this paper focuses on analyzing the phase structure of the Hayward-AdS black hole and exploring the repulsive interactions among its microscopic components.

The structure of this paper is organized as follows: Section \ref{sec2} provides a brief review of the thermodynamics of static, spherically symmetric Hayward-AdS black holes in an extended phase space.
In section \ref{sec3}, we will study the critical behavior of the Hayward-AdS black hole, and further discusses the critical exponents associated with $P$-$V$ criticality. Section \ref{sec4} focuses on the thermodynamic curvature scalar of the Hayward-AdS black hole near the critical point, analyzing its divergence behavior. Finally, Section \ref{conclusion} presents the conclusions and prospects of this study. Throughout the paper, natural units are adopted, with $c=\hbar=k_{B}=1$.


\section{Thermodynamics of the Hayward-AdS black holes}\label{sec2}

Recent advances demonstrate that infinite towers of higher-order curvature corrections to the Einstein-Hilbert action generically resolve the Schwarzschild singularity in spacetimes of dimension $D\geq 5$ \cite{Bueno:2024dgm}. These corrections underpin the framework of quasi-topological gravities \cite{Oliva:2010eb}--a class of vacuum gravitational effective field theories featuring second-order field equations in spherical symmetry and a Birkhoff theorem that guarantees the uniqueness of static solutions.
This singularity resolution via resummation of infinite curvature corrections aligns with quantum gravity predictions, and while quasi-topological gravities include Lovelock gravity \cite{Lovelock:1971yv} (which has second-order equations universally), they extend beyond the Lovelock class (limited to $n \leq D/2$):starting with a cubic theory \cite{Myers:2010ru}, extensions to quartic \cite{Dehghani:2011vu}, quintic \cite{Cisterna:2017umf} and all curvature orders \cite{Bueno:2019ycr,Bueno:2022res} have been found, all satisfying the Birkhoff theorem such that general vacuum spherically symmetric solutions are static \cite{Bueno:2024zsx}.

The action for quasi-topological gravity at every curvature order (except for $n = D/2$) and in dimensions $D\geq 5$ is given by \cite{Bueno:2024eig,Bueno:2024zsx,Hennigar:2025yqm}
\begin{equation}\label{action}
I=\int \frac{{\rm d}^D x \sqrt{|g|}}{16 \pi G_{\rm N}} \bigg[R -2 \Lambda + \sum_{n=2}^{\infty} {\alpha}_n \mathcal{Z}_n \bigg] \,,
\end{equation}
where $g$ is the metric determinant, $R$ is the Ricci scalar, ${\alpha}_n$ (length$^{2(n-1)}$) are coupling constants, and $\mathcal{Z}_n$ denotes the $n-th$ order curvature term. Here, $\mathcal{Z}_1$ corresponds to the Einstein-Hilbert term,  $\mathcal{Z}_2$ to the Gauss-Bonnet density, and higher-order terms no longer overlap with Lovelock densities. Unlike Lovelock theories, quasi-topological gravities allow infinite curvature corrections--a critical distinction enabling singularity resolution \cite{Bueno:2024dgm}.

Varying the action in Eq.(\ref{action}) yields the field equations, from which the line element of the static, spherically symmetric Hayward-AdS black hole is given by \cite{Hennigar:2025yqm}
\begin{equation}\label{eq:sss}
{\rm d}s^2 = -f(r) {\rm d} t^2 + {f(r)}^{-1}{{\rm d} r^2} + r^2 {\rm d}\Omega^2_{D-2} \,,
\end{equation}
where the metric function $f(r)$ takes the form
\begin{equation}
f(r) = 1 - \frac{r^2 \mathcal{S}(r)}{1 + \alpha \mathcal{S}(r)} \quad\quad\quad\quad \text{with} \quad\quad \mathcal{S}(r) = -\frac{1}{L^2} + \frac{m}{r^{D-1}}\,.
\end{equation}
Here, ${\rm d}\Omega^2_{D-2}$ represents the line element of a $(D-2)$-dimensional sphere, $m$ is an integration constant that is related to the Arnowitt-Deser-Misner (ADM) mass or thermodynamic mass $M$ \cite{Bueno:2024eig,Bueno:2024zsx,Hennigar:2025yqm}
\begin{align}\label{Mm}
M &= \frac{(D-2) \Omega_{D-2}m}{16 \pi G_{\rm N}}\,,
\end{align}
where $\Omega_{D-2}$ is the volume of a unit sphere $2 \pi^{(D-1)/2}/\Gamma \left[(D-1)/2 \right]$.
The event horizon of this black hole is determined by the condition $f(r_+)=0$, using which the explicit expression for mass $M$ can be written as \cite{Hennigar:2025yqm}
\begin{equation}
 M =\frac{(D-2)\Omega_{D-2} r_+^{D-1}}{16 \pi  G_{\rm N}L^2} \Big(1 + \frac{L^2}{r_+^2 - \alpha} \Big) \,.
\end{equation}
The Hawking temperature of the Hayward-AdS black hole on an event horizon is obtained as \cite{Hennigar:2025yqm}
\begin{equation}\label{THaywardAdS}
T = \frac{1}{4 \pi r_+^3} \left[(D-3) r_+^2 - \alpha (D-1) + \frac{(D-1) (r_+^2-\alpha)^2}{L^2} \right] \,,
\end{equation}
and the entropy is
\begin{equation}\label{SHaywardAdS}
S = \frac{\Omega_{D-2} r_+^{D-2}}{4 G_{\rm N}} \,\, {}_2F_1 \!\Big(2, 1 - \frac{D}{2}; 2 - \frac{D}{2}; \frac{\alpha}{r_+^2} \Big) \,.
\end{equation}
By interpreting the cosmological constant as a thermodynamic pressure as given in Eq.(\ref{PLambda}), and its conjugate quantity
 as a black hole thermodynamic volume
\begin{equation}
V = \frac{\Omega_{D-2} r_+^{D-1}}{D-1} \,,
\end{equation}
the first law of black hole thermodynamics and the corresponding Smarr formula take the form, respectively \cite{Dolan:2014vba,Hennigar:2025yqm}
\begin{eqnarray}
{\rm d}M &=& T {\rm d}S + V {\rm d P} + \Psi {\rm d} \alpha \,,\\
(D-3)M&=&(D-2)T S-2V P+2\alpha \Psi\,.
\end{eqnarray}
In the above, $\Psi=\left(\partial_\alpha M\right)_{S,P}$ is a potential conjugate to the coupling constant $\alpha$. It is worthwhile to mention that according
to Eq.(\ref{SHaywardAdS}) and black hole thermodynamic volume formula, the entropy is only a function of area/volume,
$i.e.$ $S = S(V)$. This feature of the black hole will be used in the next section.

\section{$P$-$V$ criticality of Hayward-AdS black hole}\label{sec3}

In this section, we study the $P$-$V$ phase transition and critical phenomena in the extended phase space of Hayward-AdS black holes. We first investigate the critical points and critical behavior, and then further derive the critical exponents to verify whether these exponents satisfy the scaling laws.

\subsection{$P$-$V$ phase transition and critical behavior}

By rearranging the expression in Eq. (\ref{THaywardAdS}) and combining it with Eq. (\ref{PLambda}), one can derive the thermodynamic equation of state for the $D$-dimensional spherically symmetric Hayward-AdS black hole, which is given by \cite{Hennigar:2025yqm}
\begin{equation}\label{EoSPVT}
 P=\frac{(D-2)r_+^3 T}{4 G_{\rm N} \left(r_+^2 - \alpha \right)^2} - \frac{(D-2) \left[(D-3)r_+^2 - \alpha (D-1) \right]}{16 \pi G_{\rm N} \left(r_+^2 - \alpha \right)^2} \,,
\end{equation}
where $r_+=[(D-1)V/\Omega_{D-2}]^{{1}/{(D-1)}}$. In the special case where $\alpha = 0$, the equation of state of the Hayward-AdS black hole
reduces to
\begin{equation}\label{ShwEoSPVT}
 P= \frac{(D-2) T}{4 G_{\rm N} r_+}-\frac{(D-3) (D-2)}{16\pi G_{\rm N} r_+^2} \,,
\end{equation}
which coincides exactly with the equation of state of the Schwarzschild-(A)dS black hole.\footnote{Notably, the Schwarzschild-(A)dS black hole exhibits no $P$-$V$ phase transition \cite{Dolan:2010ha,Kubiznak:2016qmn,Ma:2015llh,Hansen:2016ayo,Abdusattar:2022bpg}.}

The necessary condition for the existence of the $P$-$V$ phase transition is that \cite{Kubiznak:2012wp,Cai:2013qga,Hu:2018qsy,Abdusattar:2023pck}
\begin{equation}\label{PV}
\left(\frac{\partial P}{\partial V}\right)_{T}=\left(\frac{\partial^2 P}{\partial V^2}\right)_{T}=0\,,
\end{equation}
or equivalently
\begin{equation}\label{PVTc}
\left(\frac{\partial P}{\partial r_+}\right)_{T}=\left(\frac{\partial^2 P}{\partial r^2_+}\right)_{T}=0\,,
\end{equation}
has a critical point solution $T=T_{\rm c},\ P=P_{\rm c},\ V=V_{\rm c}$.
This system is known to exhibit a small-large black hole phase transition \cite{Hennigar:2025yqm}, where the critical point
\begin{eqnarray}\label{crit}
V_{\rm c}&=&\frac{\Omega_{D - 2}}{D-1} r_{\rm c}^{D-1}\,,\,\,\,\,\,\,r_{\rm c}=\sqrt{\frac{\Big[3 (D-1)+2 \sqrt{3} \sqrt{(D-2) D}\Big]\alpha}{D-3}}\,,\\	
T_{\rm c}&=&\frac{(D-3) \left[\sqrt{3} \sqrt{(D-2) D^5}+D \left(D (2 D-1)+\sqrt{3} \sqrt{(D-2) D}-12\right)-6 \sqrt{3} \sqrt{(D-2) D}+9\right]}{2 \pi r_{\rm c} \left[2 \sqrt{3} \sqrt{(D-2) D^5}+3 D^3-27 D-9 \sqrt{3} \sqrt{(D-2) D}+27\right]}\,, \nonumber\\
P_{\rm c}&=&\frac{\sqrt{3} (D-3)^2 (D-2)^{3/2} \sqrt{D}}{32 \pi G_{\rm N} \alpha  \left[D+\sqrt{3} \sqrt{(D-2) D}\right] \left[3 D+\sqrt{3} \sqrt{(D-2) D}-6\right]}\,.\nonumber
\end{eqnarray}
It is evident that the critical points explicitly depend on both the dimensionality $D$ of the black hole and the parameter $\alpha$.

\begin{figure}[h]
\centering
\begin{minipage}[t]{7cm}
\centering
\includegraphics[width=6.5cm,height =5.2cm]{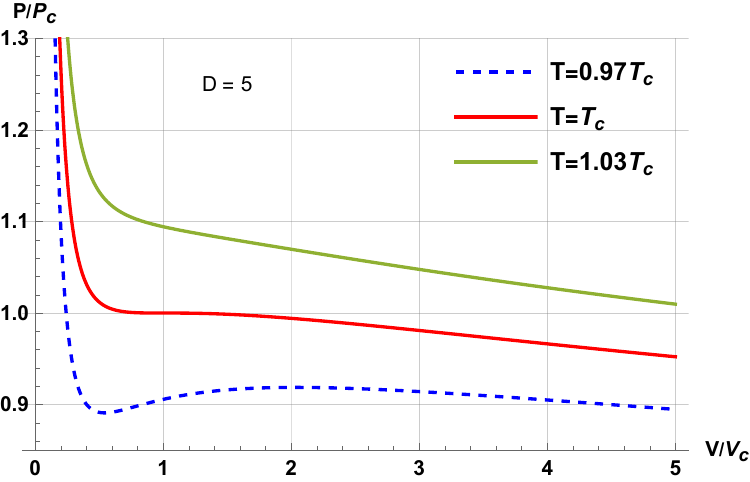}
\put(-100,-10){(a)}
\end{minipage}
\begin{minipage}[t]{7cm}
\centering
\includegraphics[width=6.5cm,height =5.2cm]{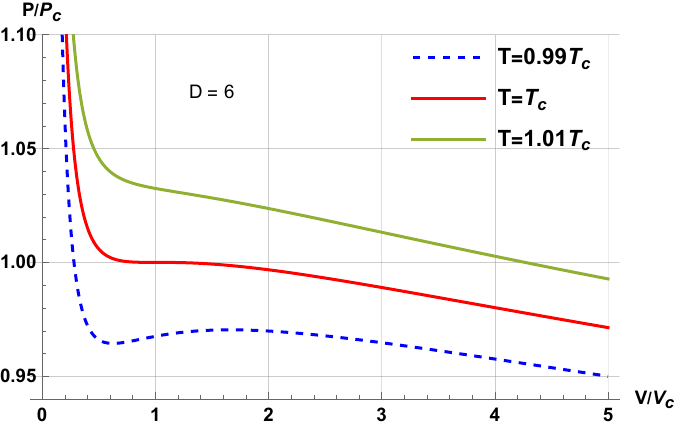}
\put(-100,-10){(b)}
\end{minipage}
\begin{minipage}[t]{7cm}
\centering
\includegraphics[width=6.5cm,height =5.2cm]{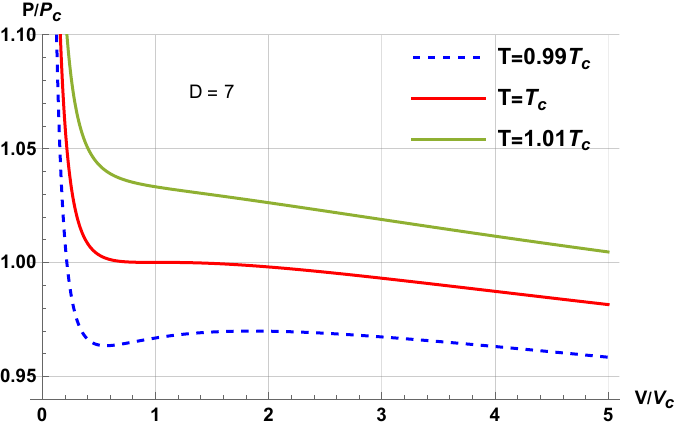}
\put(-100,-10){(c)}
\end{minipage}
\begin{minipage}[t]{7cm}
\centering
\includegraphics[width=6.5cm,height =5.2cm]{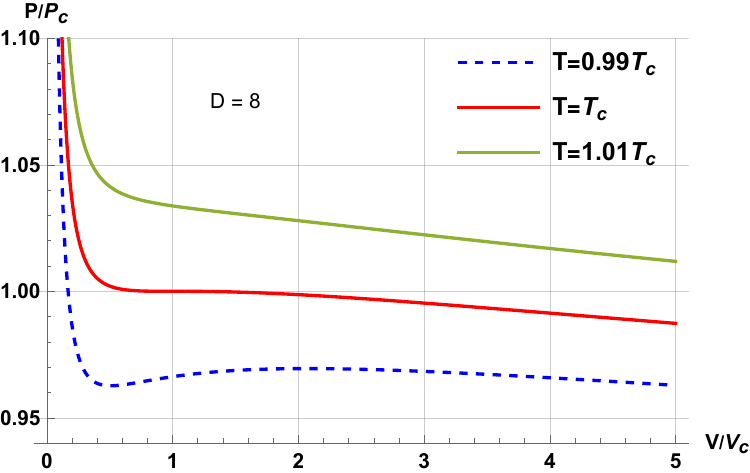}
\put(-100,-10){(d)}
\end{minipage}
\caption{$\widetilde{P}$-$\widetilde{V}$ phase diagram of Hayward-AdS black hole for different temperatures with $\alpha=0.2, G_{\rm N}=1$.}
\label{FigPV}
\end{figure}

To facilitate the clear visualization of phase diagrams, we first define dimensionless reduced pressure, reduced temperature, and reduced volume as follows
\begin{equation}\label{reduce}
\widetilde{P}\equiv\frac{P}{P_{c}}\,, \quad\quad \widetilde{V}\equiv\frac{V}{V_{c}}\,, \quad\quad \widetilde{T}\equiv\frac{T}{T_{c}}\,.
\end{equation}
Substituting these definitions into the equation of state, we obtain the reduced form of the pressure-volume relation
\begin{equation}\label{reduce}
\widetilde{P}=\frac{(D-2) \left\{\alpha  (D-1)+\left(\frac{(D-1) \widetilde{V} V_{\rm c}}{\Omega_{D-2}}\right)^{\frac{2}{D-1}} \left[3-D+4 \pi \widetilde{T} T_{\rm c} \left(\frac{(D-1) \widetilde{V} V_{\rm c}}{\Omega_{D-2}}\right)^{\frac{1}{D-1}}\right]\right\}}{16 \pi G_{\rm N} P_{\rm c} \left[\alpha -\left(\frac{(D-1) \widetilde{V} V_{\rm c}}{\Omega_{D-2} }\right)^{\frac{2}{D-1}}\right]^2} \,.
\end{equation}
The corresponding phase diagram $\widetilde{P}$-$\widetilde{V}$ is presented in Fig. \ref{FigPV}.

As depicted in Fig. \ref{FigPV}, when the temperature is below the critical value ($T<T_c$), oscillatory behavior emerges in the $\widetilde{P}$-$\widetilde{V}$ isotherms. This behavior manifests the coexistence of two distinct phases, namely small and large black holes, and thus signals a first-order phase transition between these two black hole phases analogous to the liquid-gas phase transition in a van der Waals fluid. At critical temperature ($T=T_c$), as represented by the red curves, the oscillatory feature in the plots $\widetilde{P}$-$\widetilde{V}$ vanishes, implying that a second-order phase transition occurs precisely at the critical point $(\widetilde{V}, \widetilde{P}) = (1, 1)$. For temperatures above the critical temperature ($T>T_c$), the black hole remains in a single thermodynamic phase and does not exhibit any phase transition.

\subsection{Critical exponents near the critical point}

In the following, we will calculate the critical exponents of the Hayward-AdS black hole that describe the behavior of the thermodynamic quantities near the critical point. In the usual thermodynamic system, there are four critical components, $\widetilde{\alpha}$, $\beta$, $\gamma$, and $\delta$, which are defined as follows \cite{Kubiznak:2012wp}
\begin{eqnarray}\label{abgd}
 C_V\propto |t|^{-\widetilde{\alpha}},~~~~~
 \Delta_{\widetilde{V}}=\widetilde{V}_l-\widetilde{V}_s \propto |t|^{\beta},~~~~~
 \kappa_T\propto |t|^{-\gamma},~~~~~
\widetilde{P}-1\propto \omega^{\delta},
\end{eqnarray}
where $t=\widetilde{T}-1$, $\omega=\widetilde{V}-1$.
These two reduced parameters characterize the critical behavior of the system in the vicinity of the critical point. Here, `$s$' and `$l$' are merely subscripts of $V_s$ and $V_l$, standing for `small' and `large', respectively.
Since the black hole entropy is also determined by its horizon radius or volume, $C_V$ (constant-volume heat capacity) vanishes--implying the first critical exponent $\widetilde{\alpha}=0$. To calculate the remaining three critical exponents, it is convenient to expand the equation of state given in Eq. (\ref{EoSPVT}) around the critical point
\begin{equation}\label{seriesP}
\widetilde{P}(t,\omega)=1+a_{10}t+a_{11}t\omega+a_{03}\omega^3+\mathcal{O}(t\omega^3,\omega^4)
\end{equation}
where the coefficients are
\begin{eqnarray}
 a_{10}&=&\frac{(D-2) {T_c} \left[\frac{(D-1) {V_c}}{\Omega_{D - 2}}\right]^{\frac{3}{D-1}}}{4 G_{\rm N} {P_c} \left\{\alpha -\left[\frac{(D-1) {V_c}}{\Omega_{D - 2}}\right]^{\frac{2}{D-1}}\right\}^2}>0 \,,\\
 a_{11}&=& -\frac{\left\{(D-2) {T_c} \left[\frac{(D-1) {V_c}}{\Omega_{D - 2} }\right]^{\frac{3}{D-1}} \left[3 \alpha +\left(\frac{(D-1) {V_c}}{\Omega_{D - 2} }\right)^{\frac{2}{D-1}}\right]\right\}}{4 \left\{(D-1) G_{\rm N} {P_c} \left[\left(\frac{(D-1) {V_c}}{\Omega_{D - 2}}\right)^{\frac{2}{D-1}}-\alpha \right]^3\right\}}\neq 0\,,\,\,\nonumber\\
 a_{03}&=&-\frac{\left(3 D+\sqrt{3} \sqrt{(D-2) D}-6\right) \left(3 D+2 \sqrt{3} \sqrt{(D-2) D}-3\right) \eta }{\sqrt{3} (D-1)^3 \sqrt{(D-2) D} \left(D+\sqrt{3} \sqrt{(D-2) D}\right)^4 \Theta} \,,\,\,\nonumber
\end{eqnarray}
with
\begin{eqnarray}
\eta&=&\sqrt{3} \left(11 \sqrt{(D-2) D^{11}}-13 \sqrt{(D-2) D^9}-72 \sqrt{(D-2) D^7}+117 \sqrt{(D-2) D^5}\right)\nonumber\\
&&-D \left(D (D (D (D (41-19 D)+114)-333)+189)+27 \sqrt{3} \sqrt{(D-2) D}\right)\,,\nonumber\\
\Theta&=&\left(2 \sqrt{3} \sqrt{(D-2) D^5}+3 D^3-27 D-9 \sqrt{3} \sqrt{(D-2) D}+27\right)\,,\nonumber
\end{eqnarray}
all of which are not zero. It is straightforward to show that since $\alpha$ is positive, $a_{11}$ cannot be zero and thus does not affect the critical exponent, which we demonstrate in the following discussions.
From Eq.(\ref{seriesP}), we derive $d\widetilde{P}=(a_{11}t+3a_{03}\omega^2)d\omega$, with $t$ constant along the isothermal curve. Below the critical temperature ($i.e.,$ $T<T_{c}$), the black hole system exhibits two coexisting phases: the small black hole phase and the large black hole phase. We thus denote the solutions corresponding to these phases as ($\omega_s$, $\widetilde{V}_s$) (small black hole) and ($\omega_l$, $\widetilde{V}_l$) (large black hole).

Substituting this into the Maxwell equal-area law \cite{Spallucci:2013osa,Majhi:2016txt,Bhattacharya:2017hfj,Lan:2015bia}-given by ${P}^{\ast}\left(\widetilde{V}_s-\widetilde{V}_l\right)=\int^s_l \widetilde{P}d\widetilde{V}$, we further obtain
\begin{eqnarray}\label{Maxwell E}
\int_{\omega_l}^{\omega_s} \omega d\widetilde{P} =\int_{\omega_l}^{\omega_s}\omega\frac{d\widetilde{P}}{d\omega} d\omega=\int_{\omega_l}^{\omega_s} (a_{11}t+3a_{03}\omega^2)\omega d\omega=0\,,
\end{eqnarray}
where ${P}^{\ast}$ denotes the reduced pressure corresponding to the straight line connecting the two coexisting phases. Additionally, the endpoints of the small and large black hole phases (analogous to the vapor and liquid phases in classical systems) share the same pressure $i.e.,$ ${P}^{\ast}=\widetilde{P}_l=\widetilde{P}_s$ \cite{Spallucci:2013osa,Majhi:2016txt,Bhattacharya:2017hfj,Lan:2015bia}. This implies
\begin{equation}\label{PlPs}
 a_{11}t(\omega_{l}-\omega_s)+a_{03}(\omega_{l}^3-\omega_{s}^3)=0.
\end{equation}
Solving Eqs.(\ref{Maxwell E}) and (\ref{PlPs}), yields unique non-trivial solutions (with $\omega_{l}\neq\omega_{s}$) for $t<0$. Using $\widetilde{V}_s=1+\omega_s$ and $\widetilde{V}_l=1+\omega_l$, we explicitly derive the analytical expressions for the reduced volumes of the two coexisting phases near the critical point
\begin{eqnarray}
\widetilde{V}_{s}&=&1-\sqrt{-\frac{a_{11}}{a_{03}}t}\,,\label{SV}\\
\widetilde{V}_{l}&=&1+\sqrt{-\frac{a_{11}}{a_{03}}t}\,,\label{LV}
\end{eqnarray}
where $a_{11}/a_{03}>0$ to ensure that the square root is real. From these expressions, the volume difference between the two phases is the following
\begin{equation}\label{BC}
\widetilde{V}_{l}-\widetilde{V}_{s}=2\sqrt{-\frac{a_{11}}{a_{03}}t}\propto |t|^\frac{1}{2},
\end{equation}
which gives the third critical exponent $\beta=1/2$.

The isothermal compressibility $\kappa_T$ is calculated as follows
\begin{eqnarray}
\kappa_T\propto -\left.\left(\frac{\partial \widetilde{P}}{\partial \omega}\right)_t^{-1}\right|_{\omega=0}
\propto|t|^{-1} \,,
\end{eqnarray}
yielding the critical exponent $\gamma=1$. For the critical isotherm ($t=0$), considering higher-order terms, we have $\widetilde{P}-1=a_{03}\omega^3$, which gives the final critical exponent $\delta=3$.

To conclude, we derive four critical exponents for the Hayward-AdS black hole, given by
\begin{equation}
\widetilde{\alpha}=0 \,,~~~~~~~~ \beta=\frac{1}{2}\,, ~~~~~~~~~ \gamma=1\,,~~~~~~~~~ \delta=3 \,.
 \end{equation}
These exponents are identical to those of the van der Waals liquid-gas system and thus satisfy the thermodynamic scaling laws predicted by mean field theory \cite{Kubiznak:2012wp}, as follows
\begin{eqnarray}\label{CriticalP}
&&\widetilde{\alpha}+2\beta+\gamma=2 \,, \quad\quad\quad\quad\quad\quad\quad
\widetilde{\alpha}+\beta(1+\delta)=2 \,, \nonumber \\
&&\gamma(1+\delta)=(2-\widetilde{\alpha})(\delta-1)\,,~~~~~~~~
\gamma=\beta(\delta-1) \,.
\end{eqnarray}






\section{Thermodynamic curvature via $P$-$V$ criticality}\label{sec4}

In this section, we focus on the relationship between Ruppeiner geometry (with normalized scalar curvature) and the $P$-$V$ phase transition of the spherically symmetric Hayward-AdS black hole. Ruppeiner geometry is a conceptual framework rooted in Riemannian geometry and the fluctuation theory of equilibrium thermodynamics, where entropy plays a central role. Its line element, expressed in terms of entropy, takes the form \cite{Ruppeinerb2008,Ruppeiner95}
\begin{eqnarray}\label{eq22}
{\rm d}{l^2} = g_{\mu \nu}{\rm d}{x^\mu }{\rm d}{x^\nu}\,,
\end{eqnarray}
where ${\rm d}{l^2}$ denotes the distance between two adjacent fluctuation states. The fluctuation coordinates $x = \left({U,V}\right)$ correspond to internal energy $U$ and volume $V$, and the metric tensor is given by ${g_{\mu \nu}} = - {\partial_{\mu ,\nu }}S$. For black hole systems, the first law of thermodynamics reads \cite{Ruppeinerb2008,Ruppeiner95}
\begin{eqnarray}\label{eq23}
{\rm d} S=\frac{1}{T}{\rm d}U +\frac{P}{T}{\rm d} V\,.
\end{eqnarray}
Comparing Eqs.~(\ref{eq22}) and (\ref{eq23}) and identifying the conjugate quantities of $x$ as
${y_\mu}=\partial S/\partial {x^\mu }$, leads to the following relations
\begin{subequations}\label{eq24}
\begin{align}
&{{\rm d}}\left( {\frac{1}{T}} \right) = {\left( {\frac{{{\partial^2}S}}{{\partial {U^2}}}} \right)_V}{{\rm d}}U + \left( {\frac{{{\partial^2}S}}{{\partial U\partial V}}} \right){{\rm d}}V\,, \label{eq24-1}\\
&{{\rm d}}\left({\frac{P}{T}} \right) = {\left( {\frac{{{\partial^2}S}}{{\partial {V^2}}}} \right)_U}{{\rm d}}V + \left({\frac{{{\partial^2}S}}{{\partial U\partial V}}} \right){{\rm d}}U \,. \label{eq24-2}
\end{align}
\end{subequations}
The line element in Eq.~(\ref{eq22}) can thus be rewritten as
\begin{align}
\label{eq25}
{{\rm d}}{l^2} & = - {{\rm d}}\left( {\frac{1}{T}} \right){{\rm d}}U - {{\rm d}}\left( {\frac{P}{T}} \right){{\rm d}}V
\nonumber \\
& = \frac{1}{{{T^2}}}{{\rm d}}T{{\rm d}}U - \frac{1}{T}{{\rm d}}P{{\rm d}}V + \frac{P}{{{T^2}}}{{\rm d}}T{{\rm d}}V\,.
\end{align}
Using the thermodynamic relations ${\rm d}U \!=\! {C_V}{\rm d}T + \left[ {T{{\left( {{{\partial P} \mathord{\left/ {\vphantom {{\partial P} {\partial T}}} \right.
 \kern-\nulldelimiterspace} {\partial T}}} \right)}_V} - P} \right]{\rm d}V$ and  ${\rm d} P \!=\! {\left( {{{\partial P} \mathord{\left/ {\vphantom {{\partial P} {\partial T}}} \right. \kern-\nulldelimiterspace} {\partial T}}} \right)_V}{{\rm d}}T + {\left( {{{\partial P} \mathord{\left/ {\vphantom {{\partial P} {\partial T}}} \right.
 \kern-\nulldelimiterspace} {\partial T}}} \right)_T}{{\rm d}}V$, the line element can be further expressed in the $(T,V)$ coordinate system as \cite{Wei:2019uqg,Wei:2019yvs}
\begin{eqnarray}\label{dlTV}
{\rm d}l^2&=&\dfrac{C_V}{T^2}{\rm d}T^2-\dfrac{\left(\partial_V P\right)_T}{T}{\rm d}V^2 \,,
\end{eqnarray}
which encodes information about the effective interactions between microscopic fluctuation states. For a given fluid system, the curvature scalar derived from Eq.~(\ref{dlTV}) characterizes its microstructural interactions. Notably, in Ruppeiner geometry--an information geometry framework where entropy serves as the thermodynamic potential--the black hole entropy $S$ is a function of the horizon radius $r_+$ $i.e.$, $S(r_+)$. This implies that the constant-volume heat capacity $C_V=T(\partial_T S)_V$ vanishes, leading to divergences in both the line element (\ref{dlTV}) and the curvature scalar $R$.

To resolve this divergence, a normalized curvature scalar $R_N$ is defined as \cite{Wei:2019uqg,Wei:2019yvs}
\begin{eqnarray}\label{RN}
R_{\rm N}&=&R C_V \nonumber \\
&=& \frac{{\left({{\partial_V}P} \right)_T^2 - {T^2}{{\left({{\partial_{T,V}}P} \right)}^2} + 2{T^2}{{\left({{\partial_V}P} \right)}_T}\left({{\partial _{T,T,V}}P} \right)}}{{2\left({{\partial_V}P} \right)_T^2}}\,.
\end{eqnarray}
This normalized curvature scalar provides a more robust tool for investigating the microscopic properties of black holes.

\begin{figure}[h]
\centering
\begin{minipage}[t]{7cm}
\centering
\includegraphics[width=6.5cm,height =5.2cm]{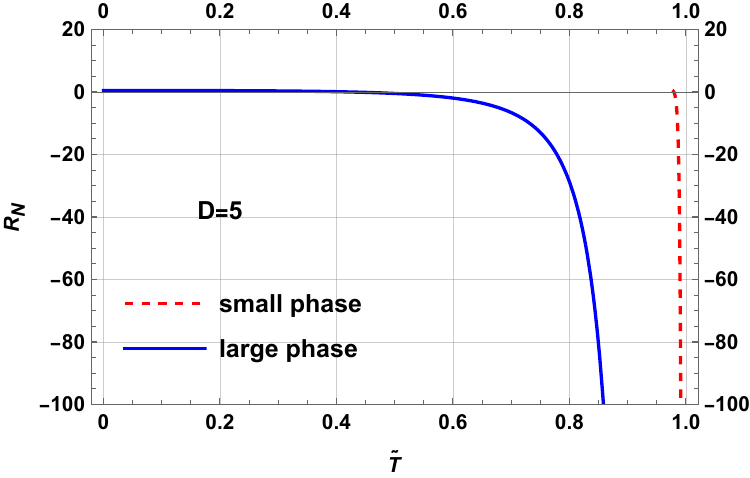}
\put(-100,-10){(a)}
\end{minipage}
\begin{minipage}[t]{7cm}
\centering
\includegraphics[width=6.5cm,height =5.2cm]{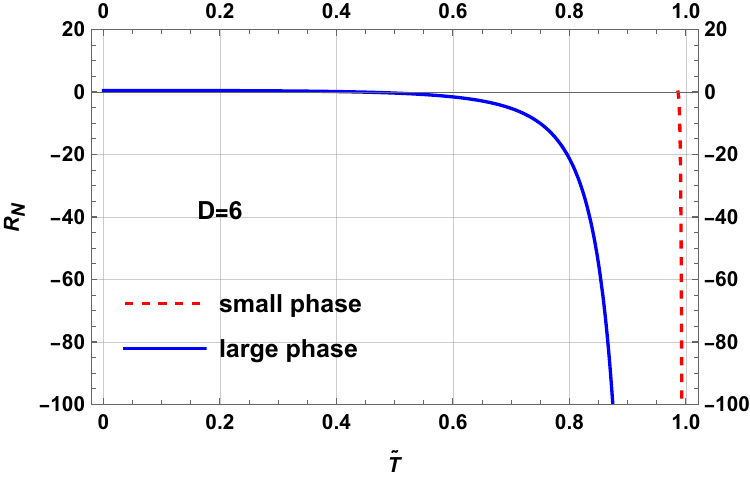}
\put(-100,-10){(b)}
\end{minipage}
\begin{minipage}[t]{7cm}
\centering
\includegraphics[width=6.5cm,height =5.2cm]{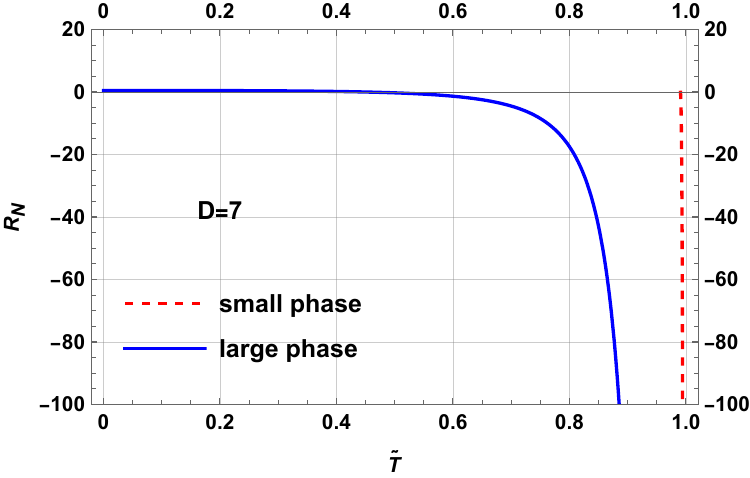}
\put(-100,-10){(c)}
\end{minipage}
\begin{minipage}[t]{7cm}
\centering
\includegraphics[width=6.5cm,height =5.2cm]{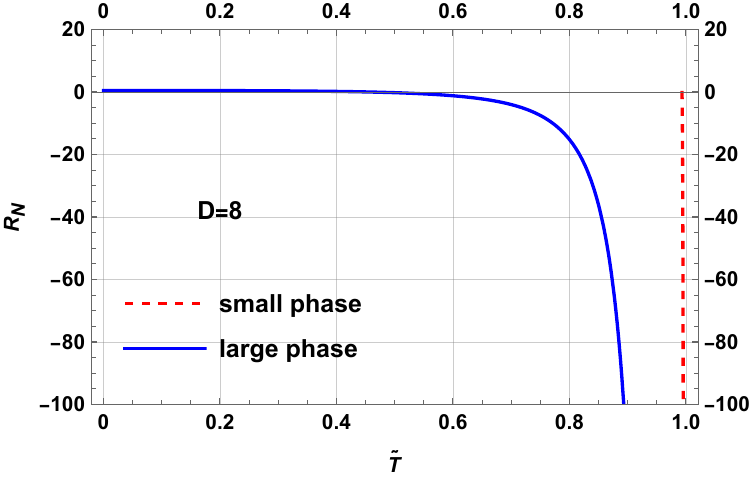}
\put(-100,-10){(d)}
\end{minipage}
\caption{{\bf Normalized Curvature Scalar $R_{\rm N}$ of Hayward-AdS Black Holes:} Behavior along the coexistence phase of small and large black holes for dimensions $D=5,6,7,8$. Here we set $\alpha=0.2, G_{\rm N}=1.$}
\label{FigRNT}
\end{figure}

After straightforward calculations, the normalized scalar curvature of the Hayward-AdS black hole expressed in terms of reduced parameters as follows
\begin{eqnarray} \label{crnn}
 R_{\rm N}=
-\frac{\left[(D-3) \zeta^{2}-\alpha (D+1)\right] \left[\alpha (D+1+12\pi \widetilde{T} {T_{\rm c}} \zeta)+(3-D+4\pi \widetilde{T} {T_{\rm c}}\zeta) \zeta^{2}\right]}{2 \left[\alpha (D+1+6 \pi \widetilde{T} {T_{\rm c}} \zeta)+\zeta^{2} (3-D+2 \pi \widetilde{T} {T_{\rm c}} \zeta)\right]^2}
\end{eqnarray}
with
\begin{eqnarray}
\zeta=\left(\frac{(D-1) \widetilde{V} {V_{\rm c}}}{\Omega_{D-2}}\right)^{\frac{1}{D-1}}\,.\nonumber
\end{eqnarray}
Note that $R_{\rm N}$ explicitly depends on the dimensions of the black hole $D$ and the parameter $\alpha$. We depict the behavior of $R_{\rm N}$ as a function of $\widetilde{T}$ in Fig.\ref{FigRNT}.

Figure \ref{FigRNT} demonstrates that at the critical point, the normalized thermodynamic curvature $R_{\rm N}$ diverges to negative infinity ($R_{\rm N} \rightarrow -\infty$) in both the small and large phases.
For $\widetilde{T}_{0}<\widetilde{T}<1$, $R_{\rm N}$ remains negative along the coexistence curve in both the small and large phases, indicating that an attractive interaction dominates.
However, along the coexistence curve for the large black hole phase at $\widetilde{T}=\widetilde{T}_{0}$, $R_{\rm N}$ can be positive as the temperature deviates from the critical temperature, implying that the microstructural interaction in the coexistence large phase of black holes undergoes a transition from attractive to repulsive during the phase transition.
Additionally, as the temperature deviates from the critical temperature, $\left\vert R_{\rm N}\right\vert $ decreases, and most of the parameter space the value of $R_{\rm N}$ is near zero, suggesting that a weak repulsive interaction dominates in Hayward-AdS black holes at low temperatures.

Near the critical point, $R_{\rm N}$ changes dramatically and tends to negative infinity, implying that the black hole microstructure changes rapidly around the temperature $\widetilde{T}_{\rm div}$
\begin{eqnarray}\label{spcurve}
 \widetilde{T}_{\rm div}=\frac{\zeta \left[(D-3) \zeta^{2} -\alpha (D+1)\right]}{2 \pi T_{\rm c} \left(3 \alpha +\zeta^{2}\right)}\,.
\end{eqnarray}

We also compute the curves where $R_{\rm N}$ changes sign, which has the following simple relation with the above divergent temperature
\begin{eqnarray}\label{sc1}
\widetilde{T}_{0}=\frac{\widetilde{T}_{\rm div}}{2}\,,
\end{eqnarray}
whose traversal indicates a change between attractive or repulsive interactions of the microstructure.
\begin{figure}[h]
\centering
\begin{minipage}[t]{7cm}
\centering
\includegraphics[width=6.5cm,height =5.2cm]{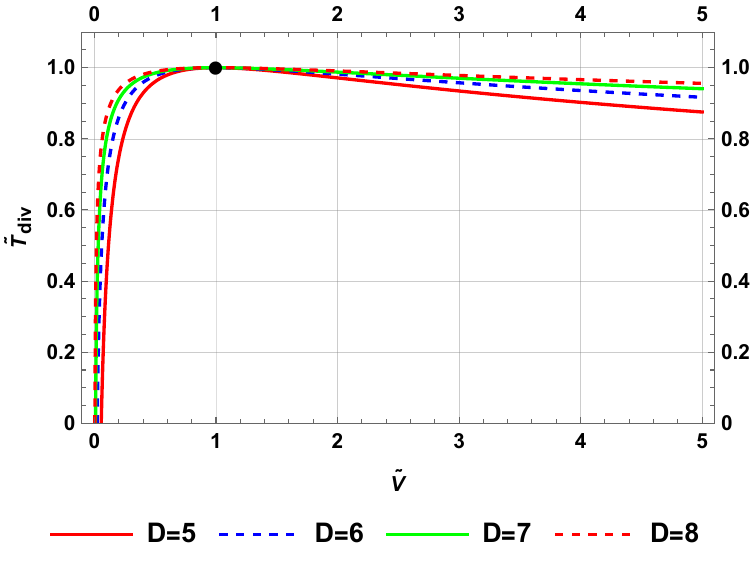}
\put(-90,-10){(a)}
\end{minipage}
\begin{minipage}[t]{7cm}
\centering
\includegraphics[width=6.5cm,height =5.2cm]{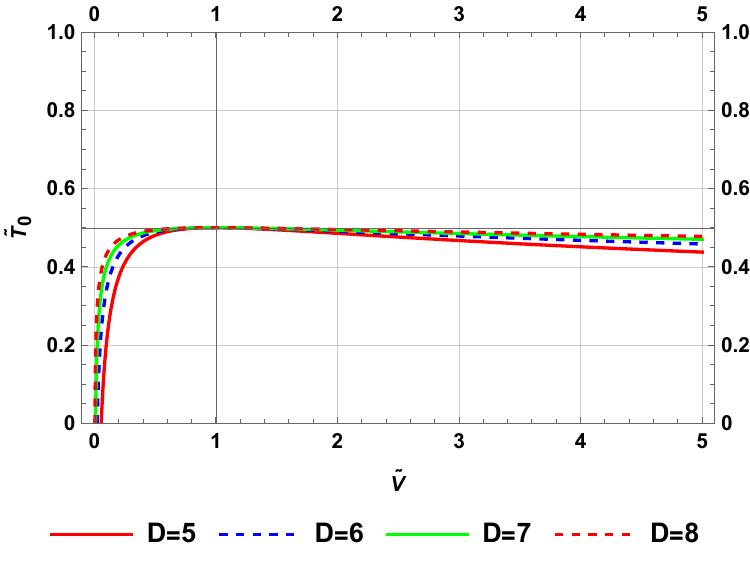}
\put(-90,-10){(b)}
\end{minipage}
\caption{{\bf Hayward-AdS black hole:}(a) The divergence curves of $R_{\rm N}$; (b) The sign changing curves of $R_{\rm N}$. Here we set $\alpha=0.2, G_{\rm N}=1.$}
\label{FigTV}
\end{figure}

From Fig.\ref{FigTV}(a), it is observed that at the critical point of the $P$-$V$ phase transition, the normalized curvature scalar $R_{\rm N}$ diverges, and the corresponding divergent temperature $\widetilde{T}_{\rm div}$ at this point shows no dependence on the spacetime dimension of the black hole (evidenced by the curves for \( D = 5, 6, 7, 8 \) converging and behaving uniformly near the critical region). Likewise, Fig.\ref{FigTV}(b) illustrates that the sign-changing temperature $\widetilde{T}_{0}$ at the critical point is also independent of the black hole's spacetime dimension, as the curves for different dimensions exhibit consistent trends around the critical point.

\section{Conclusions and Discussion}\label{conclusion}

In this study, we investigate the $P$-$V$ criticality and Ruppeiner geometry of Hayward anti-de Sitter (AdS) black holes, aiming to deepen the understanding of their thermodynamic behavior and microscopic properties. First, through thermodynamic analysis in the extended phase space, we confirm that Hayward-AdS black holes undergo distinct $P$-$V$ phase transitions and exhibit well-defined critical phenomena in the vicinity of their critical points. Notably, the four critical exponents describing these critical behaviors strictly satisfy the scaling laws predicted by mean-field theory, indicating a consistent thermodynamic framework with classical phase transition systems ($e.g.$, van der Waals fluids).

Furthermore, we use Ruppeiner geometry to investigate the microstructures of Hayward-AdS black holes--an approach that also enables probing of their thermodynamic fluctuations. We first calculated the normalized scalar curvature ($R_{\rm N}$), a quantity whose sign encodes information about the dominant microscopic interactions and which is linked to the correlation length near the critical point. Solving for $R_{\rm N}$ allowed us to identify two key temperatures: $\widetilde{T}_{\rm div}$ (the temperature at which $R_{\rm N}$ diverges) and $\widetilde{T}_{0}$ (the temperature at which $R_{\rm N}=0$). Notably, within the coexisting phase volume of small and large black holes, two such divergence temperatures exist, and these temperatures coincide exactly at the critical point.
Our analysis confirms the existence of $\widetilde{T}_{0}$ and reveals a clear transition in dominant microscopic interactions: near the critical point, attractive interactions prevail, while repulsive interactions become dominant at lower temperatures. This interaction structure is universal across most AdS black holes, and our results thus provide preliminary but valuable insights into the microstructural properties of higher-dimensional Hayward-AdS black holes along with direct understanding of the interaction nature of their microscopic constituents.


\section*{Acknowledgements}

This work is supported by National Natural Science Foundation of China (Grant No. 12465012), Kashi University high-level talent research start-up fund project (Grant No. 022024002).

\end{document}